# Exploring the Role of Interfacial Dzyaloshinskii–Moriya Interaction in Write Error Rate Anomalies of Spin-Transfer Torque Magnetic Tunnel Junctions

Prosenjit Das,[1] Md Mahadi Rajib,[1] and Jayasimha Atulasimha[1,2]

[1]Department of Mechanical and Nuclear Engineering, Virginia Commonwealth University, Richmond, Virginia, USA

[2]Department of Electrical and Computer Engineering, Virginia Commonwealth University, Richmond, Virginia, USA

**Abstract**

The performance and reliability of spin-transfer torque magnetic random-access memory (STT-MRAM) can be compromised by anomalous switching behavior, especially during high-speed operations. One such anomaly, known as the "ballooning effect" is characterized by an unexpected non-monotonic increase in the write error rate (WER) with increase in STT current at specific current pulse durations. In this study, we systematically investigate the role of the interfacial Dzyaloshinskii–Moriya interaction (DMI) on such WER anomaly using micromagnetic simulations of 20 nm and 50 nm magnetic tunnel junctions (MTJs). We show that DMI promotes incoherent magnetization reversal, prolongs the switching time and creates intermediate multidomain states that result in incomplete reversal. At high DMI values, these states persist even under large switching current densities, reproducing ballooning-like anomalies reported experimentally. In contrast, longer pulses overcome these effects by allowing the system sufficient time to reach a stable state. Our findings show that interfacial DMI can play a role in the ballooning effect and point to interfacial engineering as a practical strategy for improving the reliability of next-generation STT-MRAM.

**Introduction**

Magnetic tunnel junctions (MTJs) are the fundamental building blocks of spintronic devices and have gained significant attention due to their application in spin-transfer torque magnetic random-access memory (STT-MRAM). This nonvolatile memory technology offers high speed, high endurance, and excellent scalability, which makes MTJs a promising candidate for future memory applications [1]-[3]. However, ensuring reliable and deterministic switching in these devices remains a major challenge, particularly in the presence of thermal fluctuations and structural imperfections. As the lateral dimensions of MTJs decrease below 50 nm, switching becomes more susceptible to interfacial imperfections and stochastic thermal effects [4].

One notable concern associated with STT-MRAM is the occurrence of anomalous switching behavior known as the "ballooning effect" [5]. The ballooning effect is associated with an unexpected, non-monotonic rise in write error rate (WER) with increase in write currents (or write current densities) at certain pulse durations. This behavior is problematic for high-speed applications where short-duration pulses, such as 5 ns and 10 ns pulses are used. Previous experimental studies have reported such ballooning of WER for shorter pulses (5-10 ns) while this effect was suppressed for longer pulses (30-50 ns) [6]. Similar anomalies have also been noted in other STT-MRAM studies [7–9]. This anomalous behavior is often attributed to complex magnetization dynamics in the free layer causing non-uniform, incoherent switching patterns [7], [10-11], including thermally assisted nucleation, edge/roughness effects, and pulse-timing sensitivity [12]. These dynamics often involve formation of transient domain structures or localized switching that strongly deviate from ideal macrospin switching trajectories. Such incoherence is difficult to capture in analytical models, making micromagnetic simulations essential for understanding the physical origins of the ballooning effect [13-15]. Such an understanding would also enable the design of future STT-MRAM that is robust to such effects.

The Dzyaloshinskii-Moriya interaction (DMI) is one of the key factors that can substantially influence switching dynamics of the nanoscale MTJs. DMI is an antisymmetric exchange interaction that is significant at the interfaces between ferromagnets and heavy metals [16]-[17]. It stabilizes chiral spin structures such as skyrmions and influences domain wall motion [18]-[24]. DMI depends on the materials, interface chemistry and structure and can therefore vary across MTJs. Importantly, in HM/FM/oxide stacks the interfacial DMI is highly sensitive to oxygen chemistry and interface structure. Oxygen can amplify DMI strength and chemically absorbed oxygen on the FM surface can induce large DMI [25, 26]. Standard extraction routes and typical magnitudes are surveyed in Ref. [27]. Simulations have shown that DMI modifies the energy barrier and can lead to complex switching trajectories that deviate from the macrospin behavior.

Prior micromagnetic studies also report DMI-dependent switching thresholds and altered reversal pathways in perpendicular MTJs [28–30] In particular, DMI can change the energy profile in favor of intermediate states like multi-domains or skyrmion-like states, which can either assist or hinder the switching process depending on the pulse conditions [22].

While DMI has been studied extensively in the context of skyrmion dynamics, domain wall motion, and even degraded STT switching, its direct role in WER anomalies particularly the ballooning effect remains unexplored. Moreover, DMI can influence the thermal stability of a device by reducing the energy barrier between stable magnetization states [28]. This is particularly important in scaled MTJs where slight reduction in thermal stability can cause switching errors in the presence of thermal noise. Therefore, understanding the interplay between DMI and STT pulse-driven switching is critical for uncovering the physical origin of WER ballooning and for developing strategies to design more reliable MTJs.

In this study, we use micromagnetic simulations to establish DMI as a key physical origin of the ballooning effect. By systematically varying DMI strength across a range of current densities, we identify the conditions under which DMI amplifies ballooning anomalies and demonstrates the manner in which it reshapes the reversal trajectory, leading to persistent non-uniform intermediate states that compromise switching reliability. Our findings show that there is a direct link between interfacial DMI and WER ballooning, offering critical insight into its physical origin and underscoring interfacial engineering as a practical strategy to enhance the robustness of next-generation, high-performance STT-MRAM.

**Methodology**

To investigate how DMI affects STT switching in MTJs, we conducted micromagnetic simulations using MuMax3. Circular free layers with diameters of 20 nm and 50 nm (see Supplementary Information for details of the 50 nm geometry) and a thickness of 1 nm were modeled. The free layer is a CoFeB disk (perpendicular anisotropy), which is commonly used in perpendicular MTJ stacks. The 20 nm free layer was discretized into a 16×16×1 mesh ($\Delta x = \Delta y \approx 1.25$ nm; $\Delta z = 1$ nm) chosen to be smaller than the exchange length. For our parameters ($A = 1.3\times10^{-11}$ Jm$^{-1}$, $M_s = 1.1\times10^6$ Am$^{-1}$), the exchange length $l_{ex} = \sqrt{2A/(\mu_0 M_s^2)}$ is $\approx 4.1$ nm, confirming the cell size is well below the exchange length ($\Delta x, \Delta y \ll l_{ex}$). The parameters used in the simulations are summarized in Table 1, with effective anisotropy computed as $K_{eff} = K_{u1} - 1/2\mu_0 M_s^2$. Parameter choices follow prior CoFeB/MgO stacks [31,32]. We varied the DMI constant D from 0-5 mJ/m², which covers the current experimental upper bound (3.3 mJ/m²) in Pt/Co/MgO stacks [33]. Although the highest D reported is 3.3 mJ/m², we extended the studies to 5 mJ/m² because our aim is to understand the effect of DMI on the ballooning effect. Therefore, using the higher D limit is informative even if today's devices do not have such a large DMI.

Table 1. Parameters used in the simulation [31,32].

| Parameters | Value |
| --- | --- |
| Saturation magnetization (Ms) | $1.1\times10^6$ A m$^{-1}$ |
| Exchange constant (A) | $1.3\times10^{-11}$ J m$^{-1}$ |
| Perpendicular anisotropy constant ($K_{u1}$) | $1.12\times10^6$ J m$^{-3}$ |
| Effective perpendicular magnetic anisotropy (PMA), $K_{eff} = K_{u1} - (1/2) \mu_0 M_s^2$ | $3.6\times10^5$ J m$^{-3}$ |
| Critical DMI, $D_c = \frac{4}{\pi}\sqrt{AK_{eff}}$ | 2.753 mJ m$^{-2}$ |
| Domain wall energy, $\sigma_{wall} = 4\sqrt{AK_{eff}} - \pi D$ | 8.65 mJ m$^{-2}$ (at D=0), 5.5 mJ m$^{-2}$ (at D=1 mJ m$^{-2}$), 2.37 mJ m$^{-2}$ (at D=2 mJ m$^{-2}$), 0 mJ m$^{-2}$ (at D = 2.753 mJ m$^{-2}$), -0.77 mJ m$^{-2}$ (at D= 3 mJ m$^{-2}$) |
| Effective PMA including DMI effect *$K_{eff} = K_{u1} - (1/2) \mu_0 M_s^2 - \frac{D^2\pi^2}{16A}$ | $3.6\times10^5$ J m$^{-3}$ (at D=0), $3\times10^5$ J m$^{-3}$ (at D=1 mJ m$^{-2}$) $1.7\times10^5$ J m$^{-3}$ (at D=2 mJ m$^{-2}$), 0 J m$^{-3}$ (at D=2.753 mJ m$^{-2}$), $-7\times10^4$ J m$^{-3}$ (at D=3 mJ m$^{-2}$) NOTE: Negative PMA predicted by this formula does not imply the magnetization is necessarily in-plane as this formula is an approximation. They still have PMA. |

| Gilbert damping (α) | 0.01 |
|---|---|
| Degree of spin polarization (P) | 0.44 |
| Slonczewski parameter (Λ) | 1 |
| Secondary spin−torque parameter (ϵ′) | 0 |

\* NOTE: This formula is an approximation that applies only under the macrospin assumption, there is PMA at high DMI even though this formula estimates the PMA can fall to zero.

All simulations were conducted at room temperature (300 K) to capture thermal fluctuations, which are known to trigger stochastic switching and write errors. We applied shorter (5 ns) and longer (50 ns) STT current pulses based on previous experimental studies [5] showing WER ballooning at shorter pulses and no ballooning at longer pulses. Unless otherwise stated, Joule heating, and VCMA were neglected to isolate the effects of STT and DMI. No external magnetic field was applied, and the fixed-layer polarization vector was set along $+\hat{z}$.

The magnetization dynamics were governed by the Landau–Lifshitz–Gilbert (LLG) equation coupled with a Slonczewski spin-transfer torque term [34]-[35]:

$$\frac{\partial \vec{m}}{\partial t} = \left(\frac{-\gamma}{1+\alpha^2}\right)\left[\vec{m} \times \vec{B}_{eff} + \alpha\{\vec{m} \times (\vec{m} \times \vec{B}_{eff})\}\right] + \beta \frac{\epsilon - \alpha\epsilon'}{1+\alpha^2}\left(\vec{m} \times (\vec{m_p} \times \vec{m}) - \beta \frac{\epsilon - \alpha\epsilon'}{1+\alpha^2}(\vec{m} \times \vec{m_p})\right) \quad (1)$$

$$\beta = \frac{j_z \hbar}{M_s e d} \quad (2)$$

$$\epsilon = \frac{P\Lambda^2}{(\Lambda^2 + 1) + (\Lambda^2 - 1)(\vec{m}.\vec{m_p})} \quad (3)$$

Here, $\vec{m}$ is the normalized magnetization, γ is the gyromagnetic ratio, α is the Gilbert damping constant, and $B_{eff}$ is the total effective magnetic field. $j_z$ is the current density along z axis, d is the free layer thickness, P is the polarization of incoming spins, and $\vec{m_p}$ is fixed layer magnetization. Slonczewski parameter Λ and secondary spin torque parameter ϵ′ is related to the spacer layer.

The effective magnetic field $B_{eff}$ includes contributions from exchange, anisotropy, demagnetizing, DMI, and thermal fields [34]:

$$\vec{B}_{eff} = \vec{B}_{demag} + \vec{B}_{exchange} + \vec{B}_{anisotropy} + \vec{B}_{DMI} + \vec{B}_{thermal} \quad (4)$$

$\vec{B}_{demag}$ and $\vec{B}_{exchange}$ are effective fields due to demagnetization energy and Heisenberg exchange interaction.

$\vec{B}_{anisotropy}$ and $\vec{B}_{DMI}$ are the effective fields due to perpendicular anisotropy and DMI contribution and they are modeled as follows [34]:

$$\vec{B}_{anisotropy} = \frac{2K_{u1}}{M_s} (\vec{u} \cdot \vec{m}) \vec{u} \tag{5}$$

$$\vec{B}_{DMI} = \frac{2D}{M_s} \left( \frac{\partial m_z}{\partial x}, \frac{\partial m_z}{\partial y}, -\frac{\partial m_x}{\partial x} - \frac{\partial m_y}{\partial x} \right) \tag{6}$$

Here, $K_{u1}$ is the first order uniaxial anisotropy constant, $\vec{u}$ is the unit vector in the anisotropy direction, and D is the DMI constant. Thermal noise was modeled as a stochastic field added to each simulation run, generated according to [34]:

$$\vec{B}_{thermal} = \vec{\eta} \text{ (step)} \sqrt{\frac{2\alpha k_B T}{M_s \gamma V \Delta t}} \tag{7}$$

where $\vec{\eta}$ (step) is a vector of Gaussian-distributed random numbers, $k_B$ is the Boltzmann constant, T is the temperature, V is the cell volume, and $\Delta t$ is the simulation time step. This approach ensures proper inclusion of thermal fluctuations at 300 K.

We conducted 500 simulations each at different current densities to quantify and analyze WER. For each simulation, the initial magnetization was set to +z direction (⟨m⟩ ≈ (0.01, 0.01, 0.99)) and relaxed for 1 ns to ensure thermal equilibrium of the initial magnetization state. Following that initial relaxation period, a STT pulse was applied either for 5 ns or 50 ns based on the simulation criteria. After withdrawal of the STT pulse, we continued the simulation for 2 ns to observe whether the system settled into a stable switched or unswitched state. We considered a switching event successful if the out-of-plane component of magnetization satisfied ⟨$m_z$⟩<−0.7 after 2 ns of pulse withdrawal, which is a standard criterion used to evaluate successful switching [22]. The statistical WER was computed as the ratio of failed switching events to the total number of switching attempts.

## Results and Discussion

We began our investigation by studying the influence of DMI on the STT driven switching dynamics. Fig. 1 contrasts free layer's texture evolution with and without interfacial DMI under a 5 ns STT pulse. As shown in Fig. 1 (a), without DMI, the reversal proceeds coherently with the spins staying close to a uniform out-of-plane configuration with minor edge tilt. However, when DMI was introduced into the system, the spin orientation becomes increasingly non-uniform. As shown in Fig. 1 (b), with D = 3 mJ/m$^2$, we observe in-plane tilting and more distortion in the spin orientation. Due to these non-uniformities and additional distortion, DMI makes the reversal more incoherent and delays the switching to the target state.

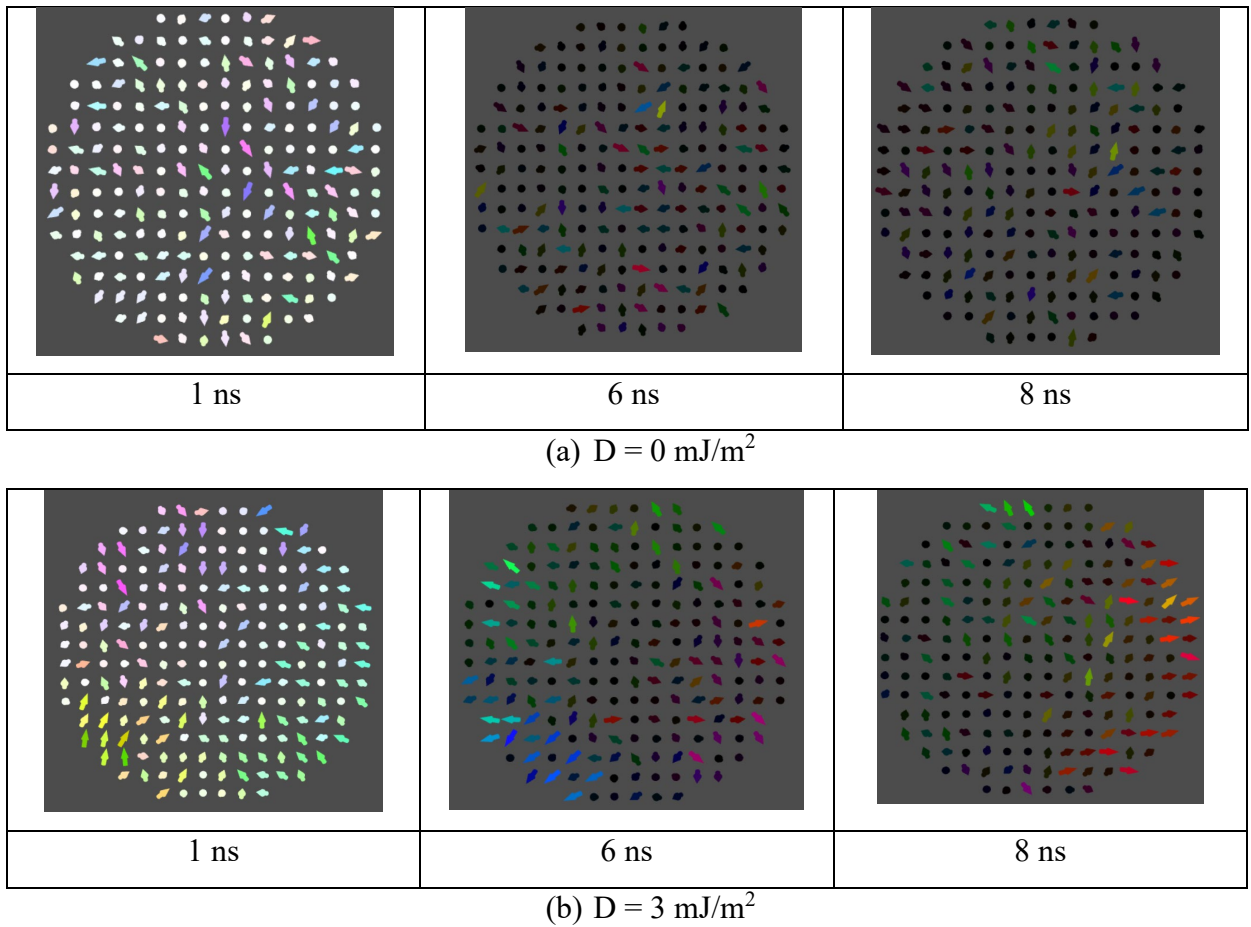

**Fig 1.** Spin configuration evolution with and without interfacial DMI under a 5 ns STT pulse at 300 K. (a) D = 0 mJ m$^{-2}$: reversal remains largely coherent with minor edge tilt. (b) D = 3 mJ m$^{-2}$: DMI induces in-plane tilting and non-uniformities in magnetization that delay reversal. Snapshots are shown at 1 ns (pre-pulse), 6 ns (immediately post-pulse), and 8 ns (late relaxation). Device: 20 nm disk, 1 nm thick CoFeB free layer. Current density = $1.2 \times 10^{11}$ A m$^{-2}$

Fig. 2 reports switching times at two different current densities: $9.6\times10^{10}$ Am$^{-2}$ and $1.2\times10^{11}$ A m$^{-2}$, comparing D = 0 mJ/m$^2$ and D=3 mJ/m$^2$. The switching time is defined as first time the out of plane component of the magnetization (m$_z$) crosses and remains below the $\langle m_z\rangle<-0.7$ target threshold.

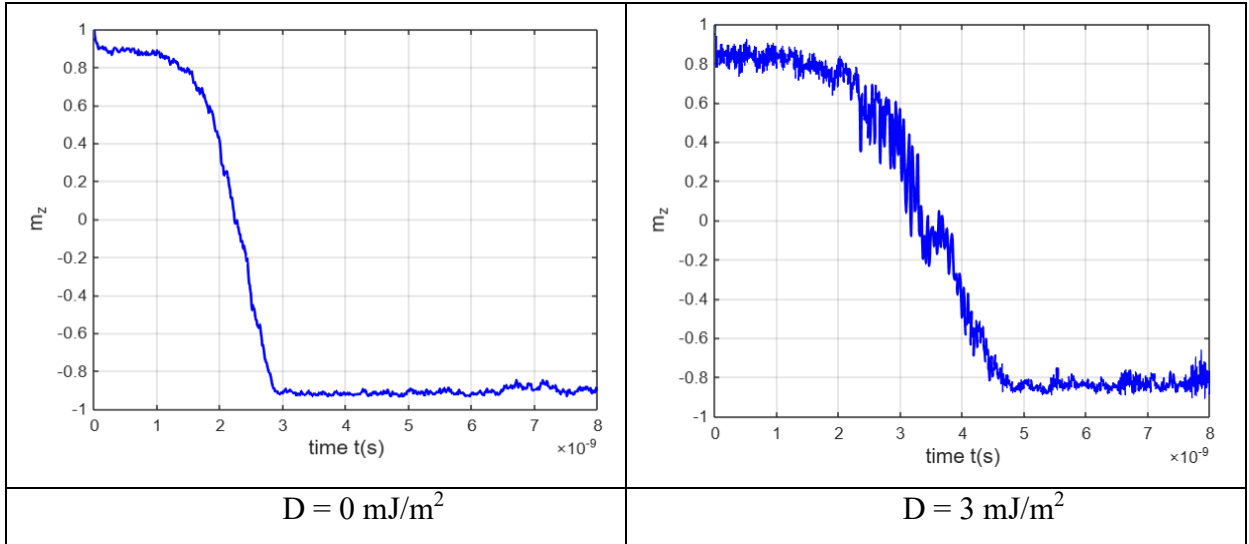

(a) $9.6\times10^{10}$ A/m$^2$

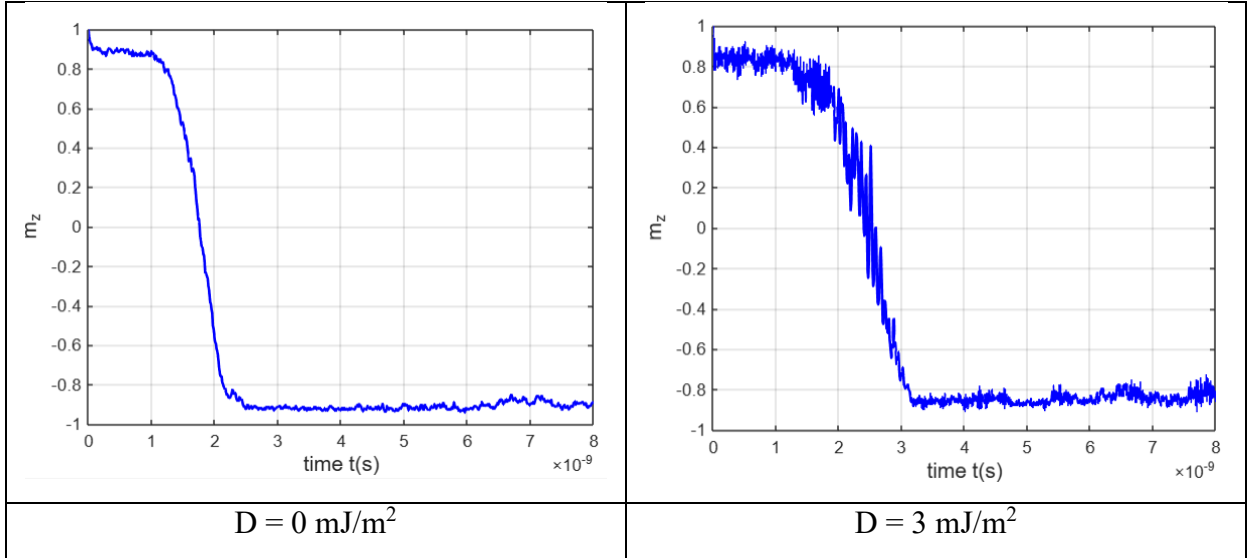

(b) $1.2\times10^{11}$ A/m$^2$

**Fig 2.** Switching-time comparison with and without DMI at two different current densities. Time evolution of $\langle m_z\rangle$ used to define switching time (first crossing that remains below the threshold) at j = $9.6\times10^{10}$ and $1.2\times10^{11}$ A m$^{-2}$ for D = 0 and D = 3 mJ m$^{-2}$. DMI slows reversal and introduces oscillations in $\langle m_z\rangle$ consistent with the spin configurations in Fig. 1.

For both current densities, DMI free reversal drops steeply to the AP state. Introducing D = 3 mJ/m$^2$ slows the reversal process and makes it non-uniform. The m$_z$ trajectory develops oscillations which are consistent with the spin configuration observed in Fig. 1(b). At higher current density (1.2×10$^{11}$Am$^{-2}$), the slope of the trajectory steepens for both cases and the DMI induced delay is reduced but not eliminated. The increased fluctuation observed in both cases for D = 3 mJ/m$^2$ explains the switching delay and also the higher write error rate in Fig 3.

Fig. 3 plots the WER on log scale versus STT current density on linear scale for different DMI constants for a 5 ns pulse at 300 K. With D = 0, WER decreases monotonically as current density increases. For D = 1-2 mJ/m$^2$, the WER curve shifts toward higher current, meaning that reaching the same WER target now needs a stronger STT drive. A more complex trend emerges at D = 3 mJ/m$^2$, where the WER curve becomes non-monotonic: after an initial decrease, WER briefly bottoms out near 10$^{-2}$, then turns upwards, and finally falls again at larger current. This non-monotonic behavior closely resembles the "ballooning" effect reported in many experimental works. At D = 3.5 mJ/m$^2$, a similar non-monotonic trend emerges, with the curve approaching a near-plateau before falling again at a large current density. For D = 4 mJ/m$^2$, the WER briefly decreases and then remains elevated over the remainder of the tested current range. At the largest DMI value, D = 5 mJ/m$^2$, the curve is nearly flat and has much high error, indicating strong DMI stabilizes the non-uniform multidomain states. The torque created by the 5 ns current drive is not enough to destabilize and reverse them into the AP state in a majority of the cases.

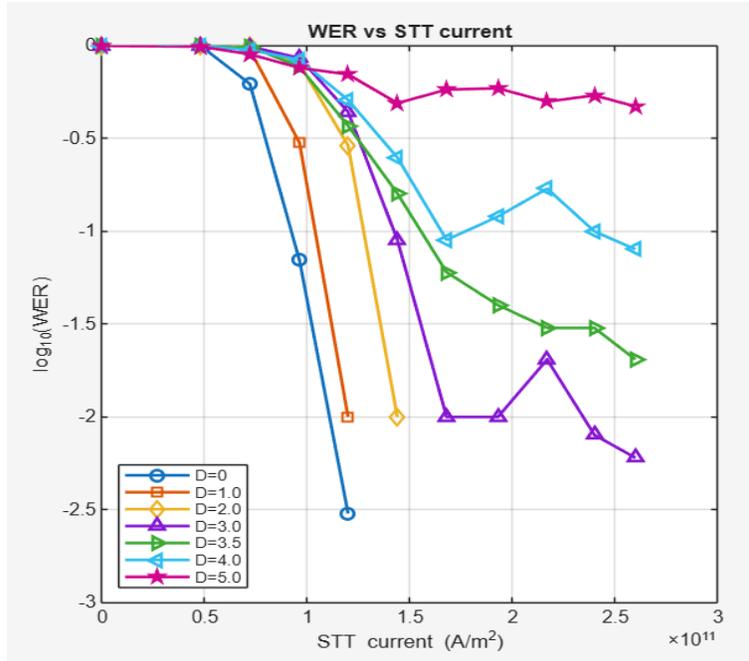

**Fig 3.** Write-error rate (WER) vs. current density for different DMI constants for 5 ns pulse at 300 K. WER is computed over 500 stochastic trials per point. D = 0 shows a monotonic decrease with current density; moderate D (≈1–2 mJ m$^{-2}$) shifts the curve to higher currents; at D = 3 mJ m$^{-2}$ and 3.5 mJ m$^{-2}$, the curve becomes non-monotonic ("ballooning"), and larger DMI (D>3.5 mJ m$^{-2}$) flattens the curve at higher WER by stabilizing non-uniform states.

Fig. 4 shows four examples of failed switching cases at D = 3 mJ/m$^2$ for a 5 ns pulse at 300 K temperature. The columns are synchronized snapshots at 1 ns (before the pulse starts), 6 ns (after the pulse ends), and 8 ns (late relaxation). In all cases a common pathway emerges. At early stage (1 ns), DMI induces in-plane tilting and the onset of non-uniformity. At post pulse (6 ns), the system has still not completed the reversal, instead a multidomain structure forms. This non-uniform configuration persists even at later stage (8 ns), so the system never reaches a uniformly switched state. Even at higher current density, this kind of non-uniform structure is created and persists beyond the pulse, leading to some failed switching even when the STT drive is strong. This micromagnetic evidence explains non-monotonic behavior as well as "ballooning" of the WER curve in Fig. 3. At D = 0 these structures do not form, so WER decreases monotonically with current density. At D= 5 mJ/m$^2$, these multi-domain states become even more favorable and appear nearly in each trial run which keeps WER high even at larger current density, as shown in supplementary figure S1.

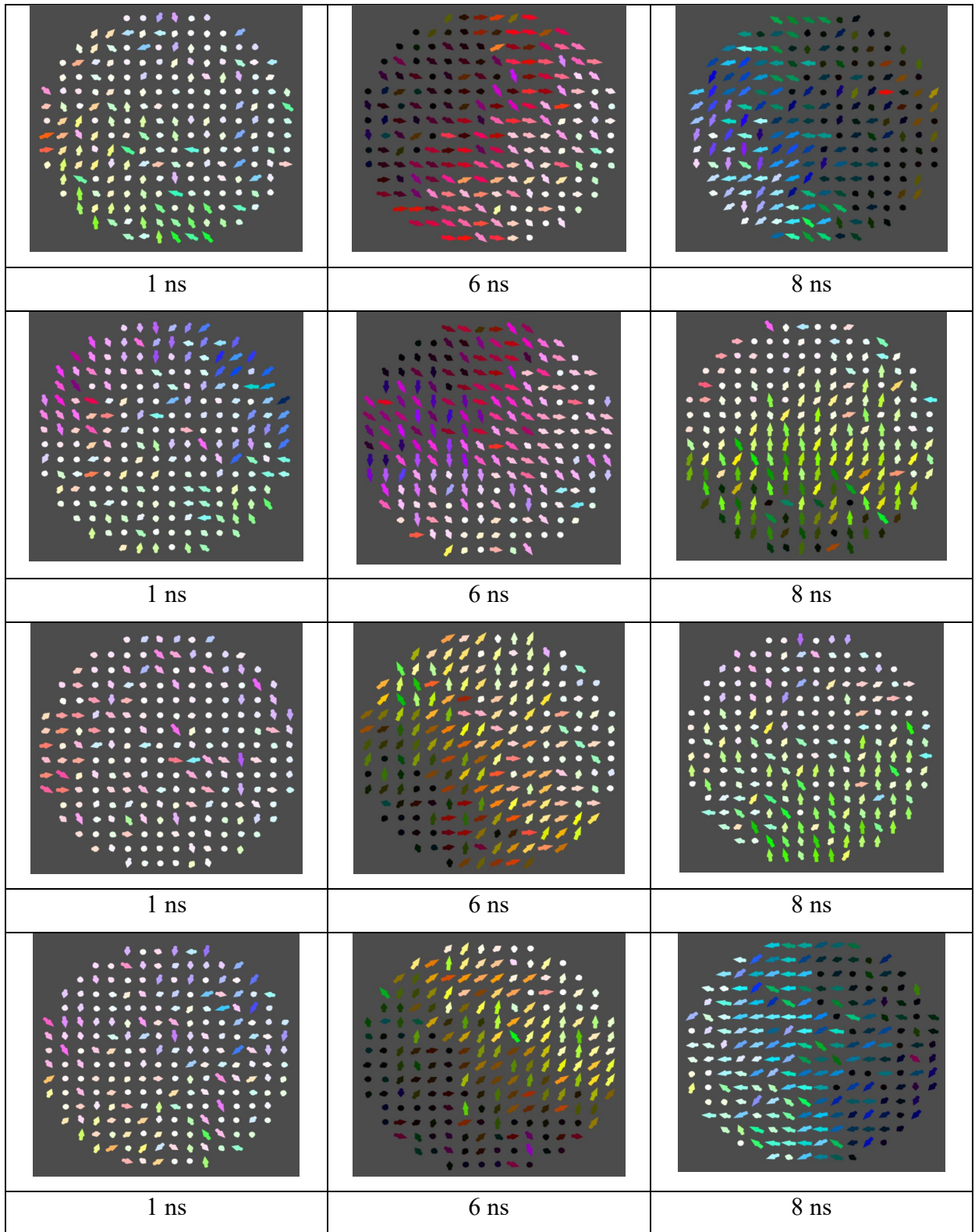

**Fig 4.** Spin evolution of four failed-switching cases at D = 3 mJ m$^{-2}$ under a 5 ns pulse at 300 K. Columns show synchronized snapshots at 1 ns, 6 ns, and 8 ns. A common pathway is observed: early in-plane tilting, post-pulse multidomain states that persist and prevent uniform reversal.

The trend observed in Fig. 3 can also be interpreted in terms of DMI-modified domain-wall energy. For our parameters, the critical DMI is $D_c$ = 2.753 mJ/m², so $D/D_c$ = {1.09, 1.27, 1.45, 1.82} for D = {3, 3.5, 4, 5} mJ/m². Near D≈$D_c$, the domain wall energy $\sigma_{wall} = 4\sqrt{AK_{eff}} - \pi D \approx 0$. Therefore, under a short STT pulse, multidomain can nucleate very easily which creates more non uniformities in the magnetization. These non-uniformities often persist post pulse as seen for D = 3 mJ/m² causing switching failure at higher current densities. As D increases further, $\sigma_{wall} < 0$, walls are energetically favored, and non-uniform multidomain states appear more frequently (see SI), which explains the high WER as D further increases.

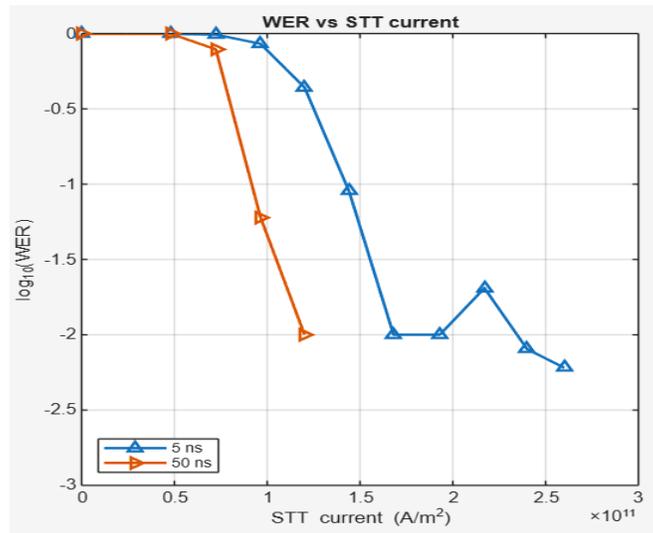

**Fig 5.** Comparing WER for short vs long pulses at D = 3 mJ m⁻². The 50 ns pulse reduces WER monotonically and at lower current density, while the 5 ns curve shows ballooning. The longer pulses provide enough time and spin transfer torque to remove non-uniform states, mitigating ballooning.

Fig. 5 compares WER curve at D = 3 mJ m⁻² for 5 ns and 50 ns pulses. At low current density, both curves have low rate of switching (large switching error). As the current density increases, the 50 ns curve drops and achieves low WER at lower write current density and the WER decreases monotonically. In contrast, the 5 ns curve shows non-monotonic behavior after an initial decrease, which resembles the ballooning effect. Thus, increasing the pulse duration mitigates the ballooning effect, indicating that longer pulse allows the non-uniform multidomain configuration to equilibrate to a more uniform state before the pulse ends.

Fig. 6 explains the reason behind observed differences in the two curves by comparing snapshots of spin configuration. As observed earlier, for the 5 ns pulse, the magnetization develops in-plane tilting, and the non-uniform structure still persists after the pulse withdrawal. For the 50 ns pulse, similar non-uniform features appear at first (at 1 ns), but by 51 ns (when the pulse is removed) there is no visible in-plane tilting and by 53 ns the magnetization largely remains in the uniformly switched state. The long pulse provides enough spin transfer torque and time to reduce the non-uniformity significantly, which results in a monotonic WER decrease and lower WER at small write currents.

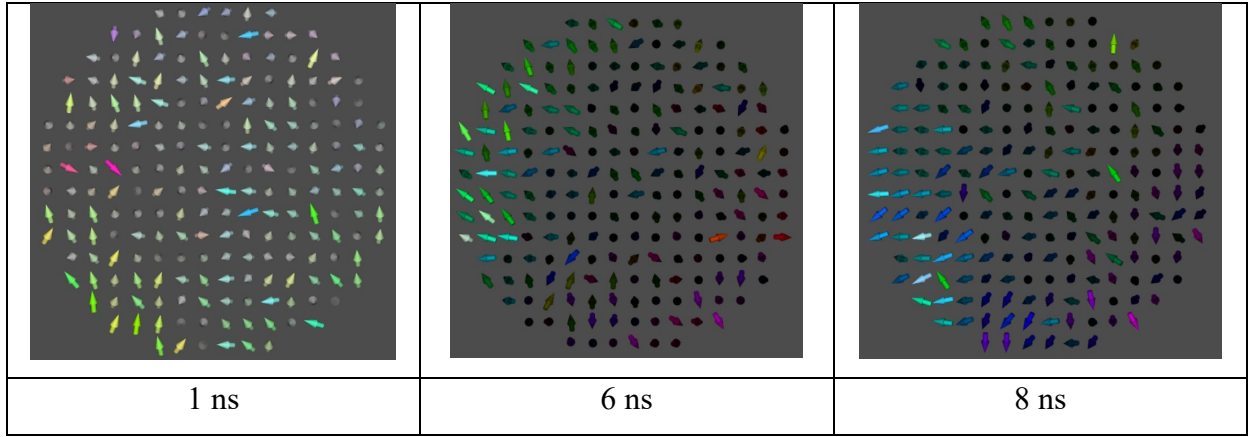

(a) For 5 ns pulse

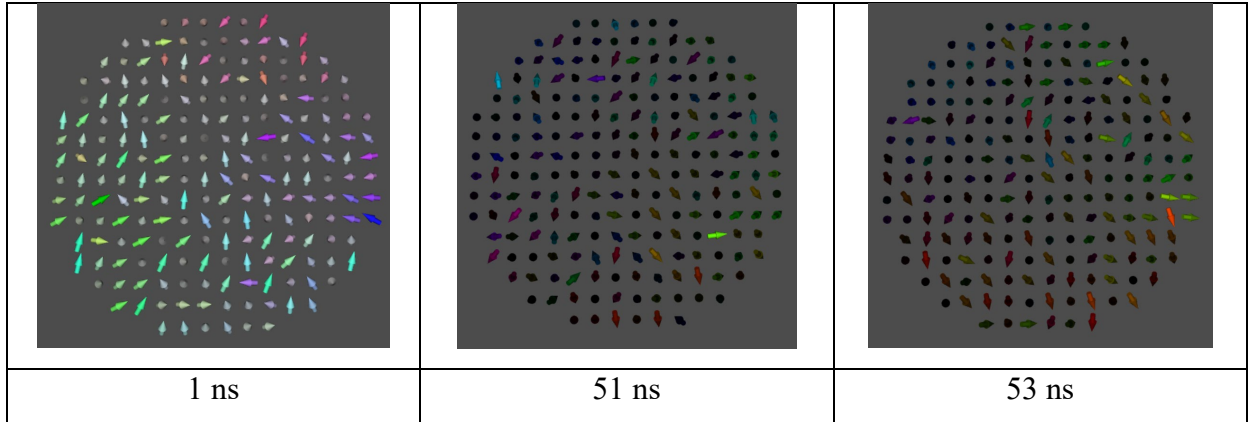

(b) For 50 ns pulse

**Fig 6.** Spin configuration evolution for 5 ns and 50 ns pulses. For 5 ns, non-uniform structures with in-plane tilting persist after the pulse; for 50 ns, similar features appear early but collapse by pulse end, and the magnetization relaxes to a uniformly switched state. Snapshots shown at 1 ns, 51 ns (pulse off), and 53 ns.

**Conclusion**

Our work establishes a link between interfacial Dzyaloshinskii–Moriya interaction (DMI) and the "ballooning" anomaly. Our results show that DMI destabilizes coherent reversal by creating in-plane tilting and multi-domain structures that persists post withdrawal of a 5 ns pulse, even at higher current densities. These non-uniform multidomain states prevent full reversal of the magnetization, which results in higher write error rate and non-monotonic ballooning behavior in WER vs. write current. In contrast, extending the pulse width to 50 ns provides sufficient torque and time to significantly reduce the non-uniform multidomain/incoherent state before the pulse ends, which restores a monotonic WER curve and eliminates the ballooning effect. Practically, these results imply that short pulse STT write reliability depends on interfacial DMI. For device design, DMI constant D should be below the onset observed in our WER-current density curve (Fig. 3). If moderate D is unavoidable, one could operate either with slightly longer pulses or at currents beyond those that lead to the non-monotonic behavior in WER vs. STT-current curve.

Our study intentionally studies STT+DMI in isolation by neglecting VCMA, joule heating and external bias field, and it focuses on circular 20 nm and 50 nm geometry (see SI for 50 nm results). As DMI induces ballooning anomaly in both 20 nm and 50 nm devices, the effect is expected to generalize across different lateral dimensions of MTJ stacks where interfacial DMI is considerable. Future work will need more extreme simulations and experiments to: (1) identify safe switching ranges in terms of current density and pulse width for additional geometries, sizes and operating temperatures, (2) examine how VCMA, joule heating and pulse shaping effect DMI-induced ballooning, (3) study the ballooning dependency on key parameters such as effective anisotropy ($k_{eff}$) and exchange stiffness (A). This work provides a foundation for future experimental and theoretical studies to further explore the ballooning phenomenon and possibly find mitigation strategies for such STT switching anomalies.

**Acknowledgement**

We acknowledge the discussion with Dr. Jonathan Z. Sun at IBM Thomas J Watson Research Center for his help with defining the problem and analyzing the results. We acknowledge support from Virginia through the Virginia VIPC CCF Commercialization grant and Virginia Commonwealth University through Virginia Commonwealth University Commercialization grant.

# Supplementary Note

# Exploring the Role of Interfacial Dzyaloshinskii–Moriya Interaction in Write Error Rate Anomalies of Spin-Transfer Torque Magnetic Tunnel Junctions

Prosenjit Das,[1] Md Mahadi Rajib,[1] and Jayasimha Atulasimha[1,2]

[1]Department of Mechanical and Nuclear Engineering, Virginia Commonwealth University, Richmond, Virginia, USA

[2]Department of Electrical and Computer Engineering, Virginia Commonwealth University, Richmond, Virginia, USA

**Spin-configuration evolution in the 20 nm geometry for large DMI (D = 4–5 mJ m⁻²):**

In the main text (Fig. 3), the WER increases sharply for $D = 4$–$5$ mJ m$^{-2}$. The underlying reason is the frequent formation and post-pulse persistence of multidomain structures during individual trials. Figure S1 shows representative failed-switching events for $D = 4$ and $5$ mJ m$^{-2}$. After a 5 ns STT pulse, the system remains partially switched, with coexisting down, up, and in-plane tilted regions. Following an additional 2 ns of relaxation, the state either remains mixed or, in some cases, fully relaxes into the initial state. Such outcomes occur more often at $D = 5$ mJ m$^{-2}$, consistent with the higher WER at that DMI.

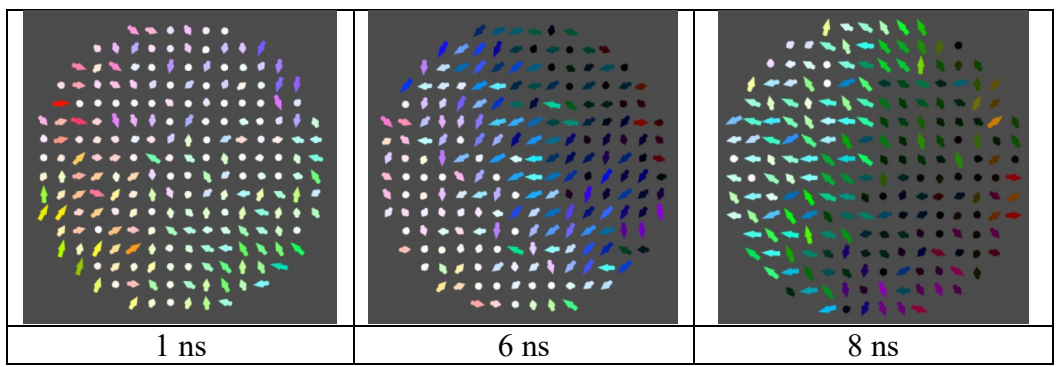

| 1 ns | 6 ns | 8 ns |

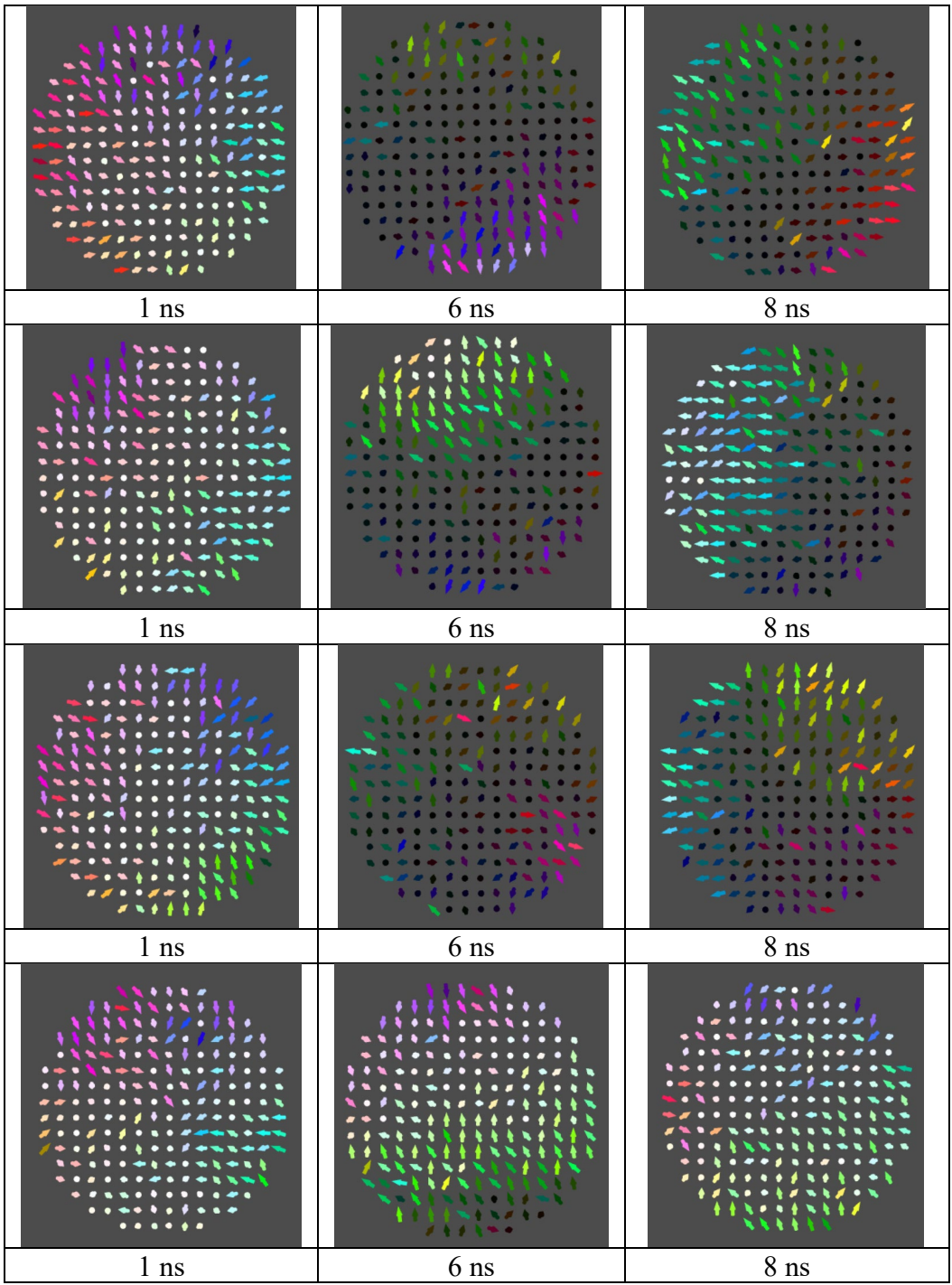

**Figure S1.** Spin-configuration evolution at 1 ns, 6 ns (end of pulse), and 8 ns (post-pulse) for D = 4 and 5 mJ m$^{-2}$ in a 20 nm device, illustrating persistent multidomain structures and failed switching.

**DMI-induced WER ballooning in a 50 nm geometry:**

To examine whether the "ballooning" non-monotonicity in write-error rate (WER) versus STT-current density persists beyond the 20 nm device, we repeated the WER study for a circular p-MTJ with 50 nm diameter using the same modeling framework and parameters as in the main text. The free layer (50 nm diameter, 1 nm thickness) was discretized into 32×32×1 mesh with cell sizes of 1.5625 nm × 1.5625 ×1 nm, which is safely below the exchange length. We varied the interfacial DMI constant D from 0 to 5 mJ m$^{-2}$ over a range of current densities. Each operating point was simulated under identical pulse timing (5 ns drive and post-pulse relaxation) to extract WER.

Figure S2 summarizes the WER versus current density for 50 nm geometry across various D. As in the 20 nm case, increasing D shifts the WER curve toward higher current density. A ballooning-like non-monotonicity emerges at D = 3 mJ m$^{-2}$. For larger D (D = 4–5 mJ m$^{-2}$), the WER stays around 100% across the tested current range due to the creation and persistence of multidomain states that the STT pulse cannot eliminate.

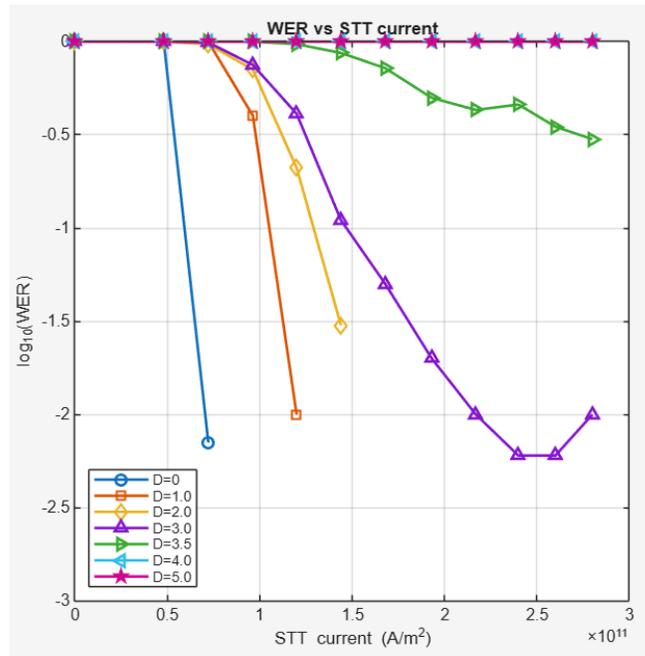

**Figure S2.** Write-error rate (WER) versus STT current density for a 50 nm p-MTJ at different DMI constants (D = 0–5 mJ m$^{-2}$). The WER curve shifts to the right with increasing D, a ballooning-like non-monotonicity appears at D = 3 mJ m$^{-2}$, while for D = 4-5 mJ m$^{-2}$ the WER ≈ 100% across the tested current range.

Failure analysis (Fig. S3) shows that at D = 3 mJ m$^{-2}$ the magnetization relaxes after the 5 ns pulse into multi domain configurations with in-plane tilting and localized up-magnetized domains within a predominantly down-magnetized background. Additional 2 ns relaxation increases the fraction of spins tipping upward, resulting in failed switching, an effect that remains observable even at higher current densities for D = 3 mJ m$^{-2}$, thereby producing the ballooning like effect as shown in Fig S2. In contrast, for D = 1–2 mJ m$^{-2}$ these mixed multidomain failures are not seen at high currents.

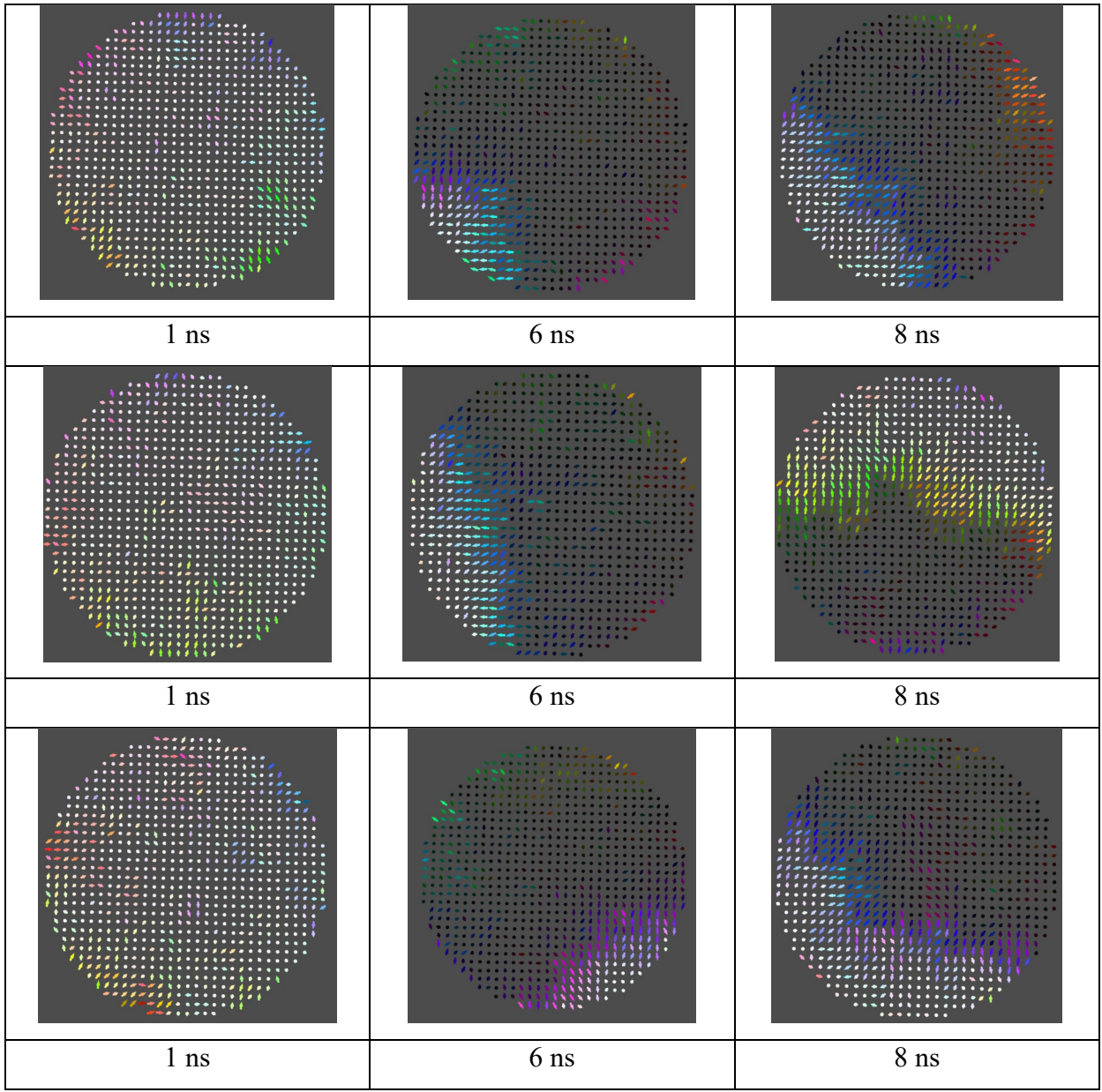

| 1 ns | 6 ns | 8 ns |
| 1 ns | 6 ns | 8 ns |
| 1 ns | 6 ns | 8 ns |

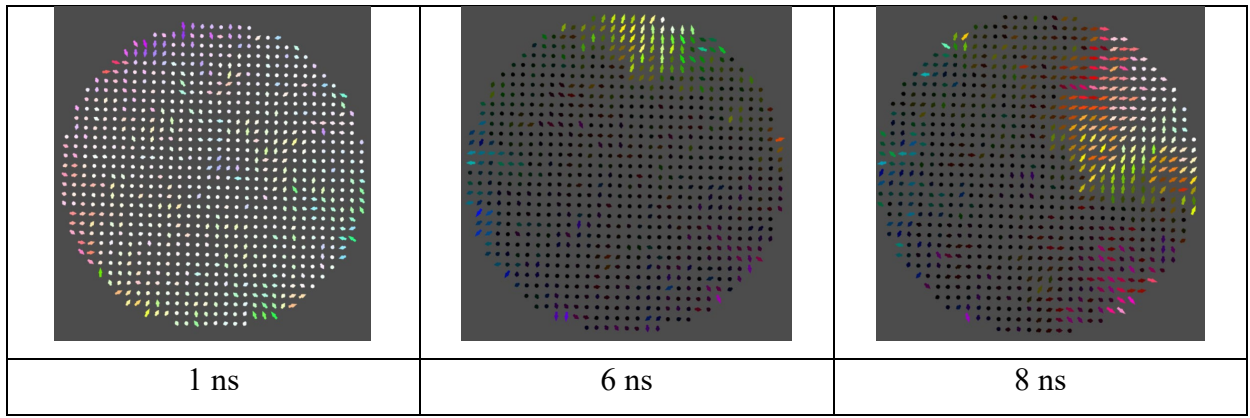

**Figure S3.** Representative failed-switching trajectories for $D = 3$ mJ m$^{-2}$ at 1 ns, 6 ns (end of STT pulse), and 8 ns (post-pulse relaxation). After pulse withdrawal, in-plane tilting and localized up-domains emerge within a largely down-magnetized background, continued relaxation increases up-domain volume and causes failure even at higher current densities.

At still higher D (4–5 mJ m$^{-2}$), the system forms robust multidomain structures across nearly all trials (Fig. S4). The applied STT amplitude in this regime is insufficient to break these structures towards the target antiparallel state, explaining the near-unity WER shown in Fig. S2.

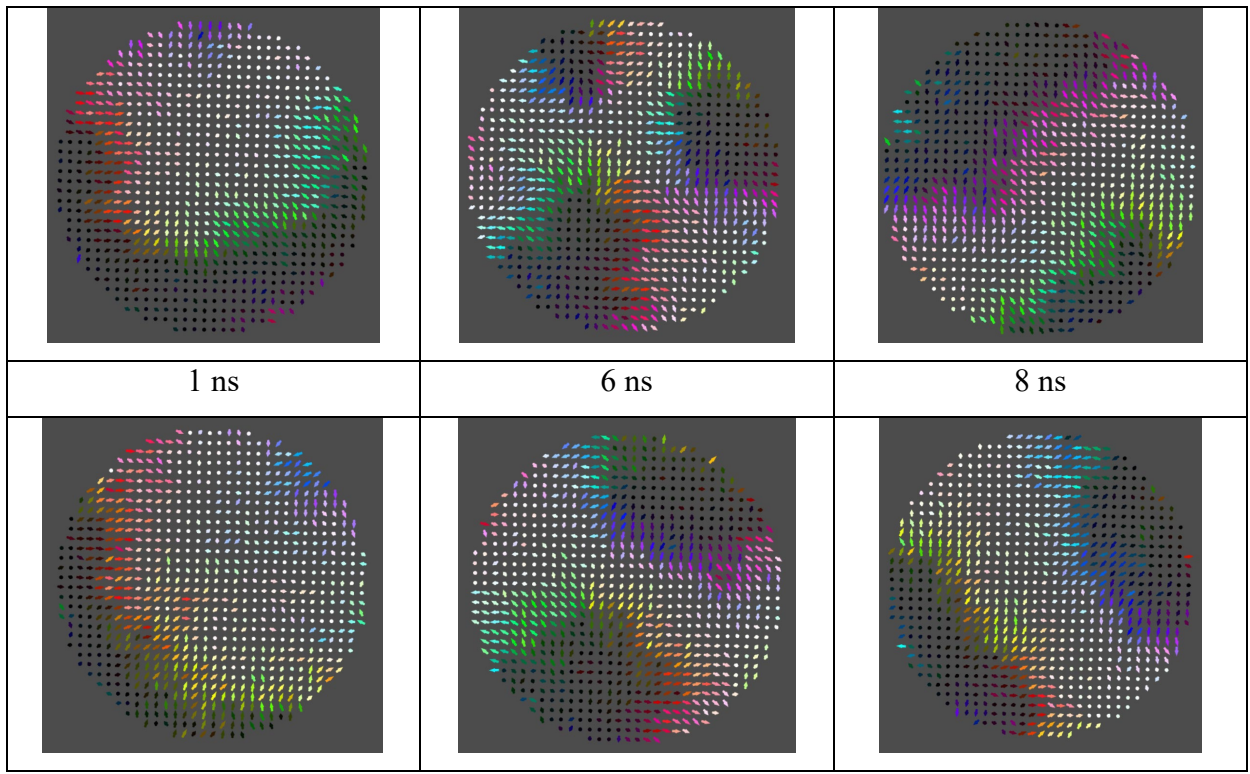

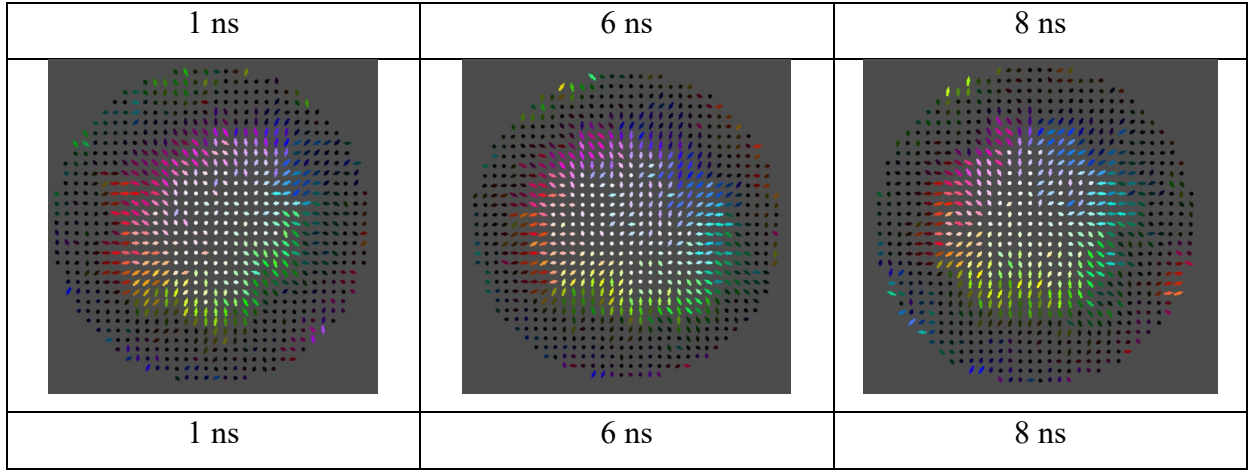

**Figure S4.** Spin-configuration evolution for D = 4–5 mJ m$^{-2}$ at 1 ns, 6 ns, and 8 ns showing robust multidomain textures that persist through and after the STT pulse, leading to near-unity WER.

The 50 nm device reproduces the qualitative WER trends reported for 20 nm: right-shifting WER curves with increasing D, a ballooning-like non-monotonicity centered near D ≈ 3 mJ m$^{-2}$, and high rate of switching failure at D = 4–5 mJ m$^{-2}$ due to multidomain formation and persistence. These observations confirm that the DMI-induced multidomain pathways that causes ballooning are not exclusive to the smallest geometry and remain relevant for larger geometries.

**DMI-Induced Switching Asymmetry**

Fig. S5 plots WER vs. STT current for (a) 20 nm and (b) 50 nm diameter geometry, comparing P→AP and AP→P reversal at D=0 and D=3 mJ/m$^2$. For D=0, the P→AP and AP→P curves are almost identical for both diameters. WER decreases almost with same rate with current for both directions, which implies that thermal barrier is direction independent. With D =3 mJ/m$^2$, the two curves show different trends. At low currents, the two curves are mostly identical, while at moderate currents, AP→P direction shows lower WER than P→AP. At higher current densities, the two curves again reconverge as strong STT drive overcomes the DMI induced preference. This directional dependency arises because interfacial DMI creates micromagnetic boundary conditions that tilt the edge of the magnetization and stabilizes domain walls of fixed chirality. When spin torque is applied it acts differently on such wall depending on the polarity. In our case, STT pushes the domain wall to favor switching for AP→P direction. Thus, the effective barrier is reduced for AP→P direction and a steeper WER can be observed. In contrast, for P→AP direction, STT opposes the preferred chirality, which creates more metastable multidomain states, resulting in higher WER. However, at sufficiently large current density, STT dominates, and the two curves converge again.

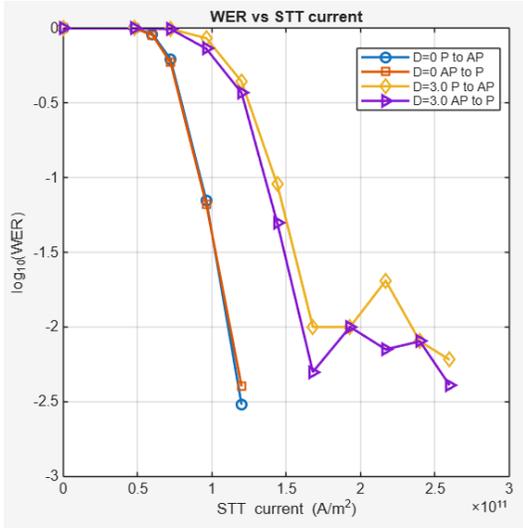
(a) 20 nm

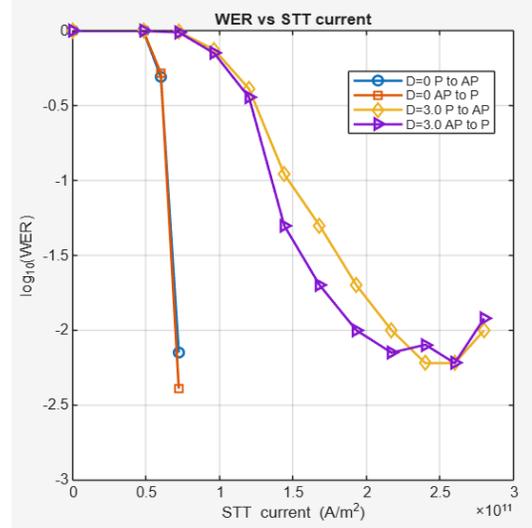
(b) 50 nm

**Fig S5.** Comparison of WER versus STT current for P→AP and AP→P reversal at D = 3 mJ m$^{-2}$ in (a) 20 nm and (b) 50 nm MTJs, showing DMI-induced asymmetry in WER.